\documentclass[a4paper,12pt]{spieman}  
\usepackage[T1]{fontenc}
\usepackage{textcomp}
\usepackage[intlimits]{amsmath}
\usepackage{amssymb,bm}
\usepackage{amsfonts}
\usepackage{setspace}
\usepackage{tocloft}

\usepackage{graphicx}
\usepackage{epstopdf}
\usepackage{setspace}
\usepackage{color}
\usepackage[final]{showkeys}
 \makeatletter

\newcommand{\rmi}{{\mathrm{i}}}
\newcommand{\rme}{{\mathrm{e}}}

\newcommand{\sign}{{\operatorname{sign\,}}}

\newtheorem{remark}{Remark}

\providecommand{\tabularnewline}{\\}

\RequirePackage{ifthen}
\makeatletter
\newcommand{\logmessage}[1]{\@latex@warning{#1}}
\makeatother
\IfFileExists{../Bibliographie/Users_Guide.txt}{
  \def\BibPath{../Bibliographie/}}{
  \IfFileExists{../../Bibliographie/Users_Guide.txt}{
    \def\BibPath{../../Bibliographie/}}{
    \IfFileExists{../../../Bibliographie/Users_Guide.txt}{
      \def\BibPath{../../../Bibliographie/}}{
        \IfFileExists{../../../../Bibliographie/Users_Guide.txt}{
          \def\BibPath{../../../../Bibliographie/}}{
            \IfFileExists{../../../../../Bibliographie/Users_Guide.txt}{
              \def\BibPath{../../../../../Bibliographie/}}{
        \logmessage{Directory 'Bibliographie' not found}
          \def\BibPath{}}}}}}

\title{Non-Equispaced Grid Sampling in Photoacoustics with a Non-Uniform FFT}

\author[a,b]{Julian~Schmid}
\author[a]{Thomas~Glatz}
\author[b]{Behrooz~Zabihian}
\author[b]{Mengyang~Liu}
\author[b]{Wolfgang~Drexler}
\author[a,c]{Otmar~Scherzer}

\affil[a]{Computational Science Center, University of Vienna, Austria}
\affil[b]{Center for Medical Physics and Biomedical Engineering, Medical University of Vienna, Austria}
\affil[c]{Johann Radon Institute for Computational and Applied Mathematics (RICAM), Austrian Academy of Sciences}

\cftpagenumbersoff{figure}
\cftpagenumbersoff{table}

\makeatother

\begin{document}

\maketitle
\begin{abstract}

To obtain the initial pressure from the collected data on a planar sensor arrangement in photoacoustic tomography, there exists an exact analytic frequency domain reconstruction formula. An efficient realization of this formula needs to cope with the evaluation of the data's Fourier transform on a non-equispaced mesh. In this paper, we use the non-uniform fast Fourier transform to handle this issue and show its feasibility in 3D experiments with real and synthetic data. This is done in comparison to the standard approach that uses linear, polynomial or nearest neighbor interpolation.
Moreover, we investigate the effect and the utility of flexible sensor location to make optimal use of a limited number of sensor points. 
The computational realization is accomplished by the use of a multi-dimensional non-uniform fast Fourier algorithm, where non-uniform data sampling is performed both in frequency and spatial domain.
Examples with synthetic and real data show that both approaches improve image quality. 

\end{abstract}

\section{Introduction}

Photoacoustic tomography is an emerging imaging
technique that combines the contrast of optical absorption with
the resolution of ultrasound images (see for instance \cite{Wan09}). 
In experiments an object is irradiated by a short-pulsed laser beam. 
Depending on the absorption properties of the material, some of the pulse
energy is absorbed and converted into heat. 
This leads to a thermoelastic expansion, which causes a pressure rise, 
resulting in an ultrasonic wave called photoacoustic signal. 
The signal is detected by an array of ultrasound transducers outside 
the object. Using this signal the initial pressure is reconstructed, 
offering a 3D image proportional to the amount of absorbed energy at each position. 
This is the imaging parameter of photoacoustics.

Common measurement setups rely on small ultrasound sensors, which are 
arranged \emph{uniformly} along simple geometries, such as planes, 
spheres, or cylinders (see for instance \cite{XuWan02a,XuWan02b,XuWan03,Wan09,Bea11}).
A non-equispaced arrangement of transducers aligned on a spherical array has already been used by \cite{XiaWanJiJia13}.
Here we investigate photoacoustic reconstructions from ultrasound signals recorded at 
\emph{not necessarily equispaced} positions on a planar surface. In other words, we use an irregular
sensor point arrangement, where sensor points are denser towards the center. This is done in order to maximize 
image quality, when the number of sensor points is the limiting factor. This approach 
can be used for dealing with the limited view problem
where deficiencies are caused by a small detection region and is motivated by the capabilities 
and requirements of our experimental setup.

For the planar arrangement of point-like detectors there exist several 
approaches for reconstruction, including numerical algorithms based on 
filtered back-projection formulas and time-reversal algorithms (see for 
instance \cite{XuWan02b,KucKun08,XuWan05,XuWan06}). 

The suggested algorithm in the present work realizes a Fourier inversion formula (see 
\eqref{eq:exact} below) using the \emph{non-uniform fast Fourier transform} (NUFFT).
This method has been designed for evaluation of Fourier transforms at non-equispaced  
points in frequency domain, or non-equispaced data points in spatial, respectively 
temporal domain. The prior is called NER-NUFFT (non-equispaced range non-uniform FFT), 
whereas the latter is called NED-NUFFT (non-equispaced data non-uniform FFT).
Both algorithms have been introduced in \cite{Fou03}. Both NUFFT methods have
proven to achieve high accuracy and simultaneously reach the computational efficiency
of conventional FFT computations on regular grids \cite{Fou03}.    

For the reconstruction we propose a novel combination of NED- and 
NER-NUFFT, which we call NEDNER-NUFFT, based on the following considerations: 
\begin{enumerate}
 \item The discretization of the analytic inversion formula \eqref{eq:exact} contains evaluations 
 at non-equidistant sample points in frequency domain. 
 \item In addition, and this comes from the motivation of this paper, we consider evaluation at 
 non-uniform sampling points.
\end{enumerate}

The first issue can be solved by a NER-NUFFT implementation: For \emph{2D} photoacoustic inversion with 
\emph{uniformly} placed sensors on a \emph{measurement line}, such an implementation has been considered 
in \cite{HalSchZan09b}. Furthermore, this method was used for biological photoacoustic imaging in \cite{SchZanHolMeyHan11}. 
In both papers the imaging was realized in 2D due to the use of integrating line detectors \cite{BurHofPalHalSch05,PalNusHalBur09}. 
In this paper we will analyze the NER-NUFFT in a \emph{3D} imaging setup with point sensors for the first time.
The second issue is solved by employing the NED-NUFFT \cite{Fou03}. Thus the
name NEDNER-NUFFT for the combined reconstruction algorithm.
%

The outline of this work is as follows:
In section \ref{sec:reconst} we outline the basics of the Fourier reconstruction approach by presenting the underlying photoacoustic model.
We state the Fourier domain reconstruction formula \eqref{eq:exact} in a continuous setting.
Moreover, we figure out two options for its discretization. 
We point out the necessity of a fast and accurate algorithm for computing the occurring discrete Fourier transforms with non-uniform
sampling points. 
In section \ref{sec:NUFFT} we briefly explain the idea behind the NUFFT.
We state the NER-NUFFT (subsection \ref{subsec:NER_NUFFT}) and NED-NUFFT (subsection \ref{subsec:NED_NUFFT}) 
formulas in the form we need it to realize the reconstruction on a non-equispaced grid.
In section \ref{sec:experimental} we introduce the \emph{3D} experimental setup.

The sections thereafter describe the realized experiments. 
In section \ref{sec:comp1_NER_NUFFT} we compare the NER-NUFFT with conventional FFT reconstruction for synthetic data in 3D.
For the real data comparisons we add a time reversal reconstruction.
Section \ref{sec:Non-Equispaced-Sensor-Placement} explains how we choose and implement 
the non-equispaced sensor placement.
In section \ref{sec:2D Comparisons} we turn to the NEDNER-NUFFT in 2D with simulated data, 
in order to test different sensor arrangements in an easily controllable environment. 
In section \ref{sec:NEDNER-realdata} we interpolate an irregular equi-steradian sensor arrangement data from experimentally acquired data-sets.
We apply our NEDNER-NUFFT approach to the non-uniform data and quantitatively compare the reconstructions to regular grid reconstructions.
We conclude with a summary of the results in section \ref{sec:results}, where we also discuss the benefits and limitations of the presented methods.

\section{Numerical Realization of a Photoacoustic Inversion Formula}
\label{sec:reconst}

Let $U \subset \mathbb{R}^d$ be an open domain in $\mathbb R^d$, 
and $\Gamma$ a $d-1$ dimensional hyperplane not intersecting $U$.
Mathematically, photoacoustic imaging consists in solving the operator equation
\[
\mathbf{Q}[f]=p|_{\Gamma\times(0,\infty)}\,,
\]
where $f$ is a function with compact support in $U$ and $\mathbf{Q}[f]$ is the trace on $\Gamma\times(0,\infty)$ of the solution of 
the equation

\begin{equation*}
 \begin{aligned}
  \partial_{tt} p - \Delta p &=0 \text{ in } \mathbb R^d \times (0,\infty)\,,\\
  p(\cdot,0) &=f(\cdot)  \text{ in } \mathbb R^d\,,\\
  \partial_t p(\cdot,0) &=0  \text{ in } \mathbb R^d\;.
 \end{aligned}
\end{equation*}
In other words, the photoacoustic imaging problem consists in identifying the initial source 
$f$ from measurement data $g=p|_{\Gamma\times(0,\infty)}$. 

An explicit inversion formula for $Q$ in terms of the Fourier transforms of $f$ and $g:=\mathbf Q[f]$ has been first formulated by \cite{NorLin81} 
and introduced to photoacoustics by \cite{KoeFreBebWeb01}. 
Let $(\bm x,y)\in \mathbb R^{d-1}\times\mathbb R^+$. 
Assume without loss of generality (by choice of proper basis) that $\Gamma$ is the hyperplane described by $y=0$.
Then the reconstruction reads as follows:
\begin{align}
\label{eq:exact}
\mathbf{F}[f]\left(\bm{K}\right)=
\frac{2K_{y}}{\kappa\left(\bm{K}\right)}\mathbf{F}[\mathbf{Q}f]\left(\bm K_{\bm x},\kappa\left(\bm{K}\right)\right).
\end{align}
where $\mathbf{F}$ denotes the $d$-dimensional Fourier transform:
\begin{align*}
\mathbf{F}[f]\left(\bm{K}\right):=\frac1{(2\pi)^{n/2}}\int\limits_{\mathbb{R}^{d}}\rme^{-\rmi\bm{K}\cdot(\bm{x},y)}f(\bm{x})\mathrm{d}\bm{x}\,,
\end{align*}
and
\begin{align*}
\kappa\left(\bm{K}\right)&=\mathrm{sign}\left(K_{y}\right)\sqrt{\bm{K_{x}}^{2}+K_{y}^{2}}\,,\\
\bm{K}&=(\bm {K_{x}},K_{y})\;.
\end{align*}
Here, the variables $\bm x,\bm {K_x}$ are in $\mathbb R^{d-1}$, whereas $y,K_y\in\mathbb R$.

For the numerical realization these three steps have to be realized in discrete form:
We denote evaluations of a function $\varphi$ at sampling points 
$(\bm x_m,y_n)\in (-X/2,X/2)^{d-1}\times(0,Y)$ by 
\begin{equation}\label{eq:eval_general}
\varphi_{m,n}:=\varphi(\bm x_m,y_n)\;.
\end{equation}
For convenience, we will modify this notation in case of evaluations 
on an equispaced Cartesian grid. We define the $d$-dimensional grid
\begin{align*}
\mathbf G_x\times \mathrm G_y:=\{-N_x/2,\dots,N_x/2-1\}^{d-1}\times \{0,\dots,N_y-1\}\,,
\end{align*}
and assume our sampling points to be located on $\bm m\Delta_x,n \Delta_y$, 
where
\[
(\bm m,n)\in \mathbf G_x\times \mathrm G_y\,,
\]
and write
\begin{equation}\label{eq:eval_equi}
\varphi_{\bm m,n}=\varphi(\bm m\Delta_x,n\Delta_y)\,,
\end{equation}
where $\Delta_x:=X/N_x$ resp.~$\Delta_y:=Y/N_y$ are the occurring step sizes.

In frequency domain, we have to sample symmetrically with respect to $K_y$. Therefore, we also introduce the interval
\[\mathrm G_{K_y}:=\{-N_y/2,\dots,N_y/2-1\}.\]
Since we will have to deal with evaluations that are partially in-grid, partially not necessarily in-grid,
we will also use combinations of \eqref{eq:eval_general} and \eqref{eq:eval_equi}.
In this paper, we will make use of discretizations of the source function $f$, 
the data function $g$ and their Fourier transforms $\hat f$ resp.~$\hat g$.

Let in the following 
\begin{align*}
\hat{f}_{\bm j,l}\,=\, \sum_{(\bm m,n)\in \mathbf G_x\times \mathrm G_y}f_{\bm m,n}\rme^{-2\pi\rmi(\bm j\cdot\bm m+ln)/(N_x^{d-1}N_y)}
\end{align*}
denote the $d$-dimensional discrete Fourier transform with respect to space and time. 
By discretizing formula \eqref{eq:exact} via Riemann sums it follows
\begin{align}\label{eq:equiv_rec}
\begin{aligned}
\hat{f}_{\bm j,l}\,\approx &\,\frac{2l}{\kappa_{\bm j,l}}\sum\limits_{n\in \mathrm G_y}\rme^{-2\pi \rmi\,\kappa_{\bm j,l}n/N_y}\\
&\cdot\underset{\bm m\,\in\,\mathbf G_x}{\sum} \rme^{-2\pi \rmi(\bm j\cdot\bm m + l n)/N_x^{d-1}}g_{\bm m,n}\,,
\end{aligned}
\end{align}
where
\begin{align*}
\hfill\kappa_{\bm j,l}&=\sign (l)\sqrt{\bm j^2+l^2}\,,\\
(\bm j,l)&\in\mathbf G_x\times\mathrm G_{K_y}\;.
\end{align*}
This is the formula from \cite{HalSchZan09b}.
\begin{remark}
Note that we use the interval notation for the integer multi-indices for notational convenience.
Moreover, we also choose the length of the Fourier transforms to be equal to $N_x$ in the first $d-1$ dimensions, 
respectively.  
This could be generalized without changes in practice.  
\end{remark}
Now, we assume to sample $g$ at $M$, not necessarily uniform, points 
$\bm x_m\in (-X/2,X/2)^{d-1}$:
Then, 
\begin{align}\label{eq:disc_rec}
\begin{aligned}
\hat{f}_{\bm j,l}\,\approx &\, \frac{2l}{\kappa_{\bm j,l}}\underset{n\in\mathrm G_y}{\sum}e^{-2\pi \rmi\kappa_{\bm j,l}n/N_y}\\
&\cdot\underset{m=1}{\overset{M}{\sum}}\; \frac{h_m}{\Delta_x^{d-1}} e^{-2\pi \rmi(\bm j\cdot \bm x_m)/M}g_{m,n}\;.
\end{aligned}
\end{align}
The term $h_m$ represents the area of the detector surface around $\bm x_m$ 
and has to fulfill $\underset{m=1}{\overset{M}{\sum}}h_m=(N_x\Delta_x)^{d-1}=X^{d-1}$.
Note that the original formula \eqref{eq:equiv_rec} can be received from \eqref{eq:disc_rec} by choosing $\{\bm x_m\}$
to contain all points on the grid $\Delta_x \mathbf G_x$.
  
Formula \eqref{eq:disc_rec} can be interpreted as follows: Once we have computed
the Fourier transform of the data and 
evaluated the Fourier transform at non-equidistant points with respect to the third coordinate, 
we obtain the (standard, equispaced) Fourier coefficients of $f$. 
The image can then be obtained by applying standard FFT techniques.

The straightforward evaluation of the sums on the right hand side of \eqref{eq:disc_rec} would lead to a computational 
complexity of order $N_y^2\times M^2$. Usually this is improved by the use of FFT methods, which have the drawback 
that they need both the data and evaluation grid to be equispaced in each coordinate.   
This means that if we want to compute \eqref{eq:disc_rec} efficiently, 
we have to interpolate both in domain- and frequency space. 
A simple way of doing that is by using polynomial interpolation.
It is used for photoacoustic reconstruction purposes for instance in the
\emph{k-wave} toolbox for Matlab \cite{TreCox10}.
Unfortunately, this kind of interpolation seems to be sub-optimal for Fourier-interpolation 
with respect to both accuracy and computational costs \cite{Fou03, XuFenWan02} 

A regularized inverse k-space interpolation has already been shown to yield
better reconstruction results \cite{JaeSchuGerKitFre07}. 
The superiority of applying the NUFFT, compared to linear interpolation,
has been shown theoretically and computationally by \cite{HalSchZan09b}.

\section{The non-uniform fast Fourier transform (NUFFT)}
\label{sec:NUFFT}
 
This section is devoted to the brief explanation 
of the theory and the applicability of the non-uniform Fourier transform, 
where we explain both the NER-NUFFT (subsection \ref{subsec:NER_NUFFT}) 
and the NED-NUFFT (subsection \ref{subsec:NED_NUFFT}) in the form
(and spatial dimensions) we utilize them afterwards.

The NEDNER-NUFFT algorithm used for implementing \eqref{eq:disc_rec} essentially (up to scaling factors) consists of the following steps:
\begin{enumerate}
\item
Compute a $d-1$ dimensional NED-NUFFT in the $\bm x$-coordinates due to our detector placement.
\item
Compute a one-dimensional NER-NUFFT in the $K_y$-coordinate as indicated by the reconstruction formula \eqref{eq:disc_rec}.
\item
Compute an equispaced $d$-dim inverse FFT to obtain a $d$ dimensional picture of the initial pressure distribution.
\end{enumerate}
\subsection{The non-equispaced range (NER-NUFFT) case}\label{subsec:NER_NUFFT}
With the NER-NUFFT (non-equispaced range -- non-uniform
FFT) it is possible to efficiently evaluate the discrete Fourier transform
at non-equispaced positions in frequency domain. 

To this end, we introduce the one dimensional discrete Fourier transform, evaluated
at non-equispaced grid points $\kappa_l\in\mathbb R$:
\begin{align}\label{eq:NUDFT1}
\hat{\varphi}_{l}=\underset{n\in \mathrm G_y}{\sum}\varphi_{n}\rme^{-2\pi \rmi\kappa_{l}n/N},\quad l=1,\ldots,M.
\end{align}
In order to find an efficient algorithm for evaluation of \eqref{eq:NUDFT1}, 
we use a window function $\Psi$, an oversampling factor $c>1$ and
a parameter $c<\alpha<\pi(2c-1)$ that satisfy:
\label{en:psi} 
\begin{enumerate}
\item $\Psi$ is continuous inside some finite interval $[-\alpha,\alpha]$
and has its support in this interval and 
\item $\Psi$ is positive in the interval $[-\pi,\pi]$. 
\end{enumerate}
Then (see \cite{Fou03,HalSchZan09b}) we have the following representation 
for the Fourier modes occurring in (\ref{eq:NUDFT1}): 
\begin{align}\label{eq:NUFFT1}
\begin{aligned}
e^{-\rmi x\theta}\,=\,\frac{c}{\sqrt{2\pi}\Psi(\theta)}\sum\limits_{k\in\mathbb{Z}}\hat{\Psi}(x-k/c)\rme^{-\rmi k\theta/c},\; |\theta|\leq\pi\;.
\end{aligned}
\end{align}
By assumption, both $\Psi$ and $\hat \Psi$ are concentrated around $0$. So we approximate the sum over all $k\in\mathbb Z$ 
by the sum over the $2K$ integers $k$ that are closest to $\kappa_l+k$. 
By choosing $\theta=2\pi n/N-\pi$ and inserting \eqref{eq:NUFFT1} in \eqref{eq:NUDFT1}, we obtain
\begin{align}\label{eq:NER_NUFFT}
\begin{aligned}
\hat\varphi_{l}\,&\approx   \,
\sum\limits_{k=-K+1}^K \hat{\Psi}_{l,k}\sum\limits_{n\in \mathrm G_y}
\frac{\varphi_{n}}{\Psi_{n}}e^{-2\pi i ln/cN}\,,\\
l \,& = \, 1,\l \dots,M\;.
\end{aligned}
\end{align}
Here $K$ denotes the interpolation length and 
\begin{align}\label{eq:PSI}
\begin{aligned}
 \Psi_n  &:=\Psi(2\pi n/N_y-\pi)\,,\\
 \hat{\Psi}_{l,k}  &:=\frac{c}{\sqrt{2\pi}}\,\rme^{-\rmi\pi(\kappa_l-(\mu_{l,k}))}\hat{\Psi}(\kappa_{l}-(\mu_{l,k}))\,,
\end{aligned}
\end{align}
where $\mu_{l,k}$ is the nearest integer (i.e. the nearest
equispaced grid point) to $\kappa_{l}+k$.

The choice of $\Psi$ is made in accordance with the assumptions
above, so we need $\Psi$ to have compact support. Furthermore, to
make the approximation in (\ref{eq:NER_NUFFT}) reasonable, its Fourier
transform $\hat{\Psi}$ needs to be concentrated as much as possible
in $[-K,K]$. In practice, a common choice for $\Psi$ is the Kaiser-Bessel
function, which fulfills the needed conditions, and its Fourier transform
is analytically computable.

\subsection{The non-equispaced data (NED-NUFFT) case}\label{subsec:NED_NUFFT}

A second major aim of the present work is to handle data measured
at non-equispaced acquisition points $\bm x_{m}$ in an efficient and accurate
way. Therefore we introduce the non-equispaced data, $d-1$ dimensional DFT
\begin{align}\label{eq:NUDFT2}
\begin{aligned}
\hat{\varphi}_{\bm j}&=\underset{m=1}{\overset{M}{\sum}}\varphi_{m}\rme^{-2\pi \rmi (\bm j\cdot \bm x_{m})/N}\,,\\
\bm j&\in\mathbf G_x\;.
\end{aligned}
\end{align}
The theory for the NED-NUFFT is largely analogous to the NER-NUFFT \cite{Fou03} 
as described in Subsection \ref{subsec:NER_NUFFT}. 
The representation \eqref{eq:NUFFT1} is here used for each entry of $\bm j$ and inserted 
(with now setting $\theta=2\pi n/N$) into formula \eqref{eq:NUDFT2}, which leads to
\begin{align}\label{eq:NED_NUFFT}
\begin{aligned}
\hat\varphi_{\bm j}\,\approx &  
\,\frac1{\Psi_{\bm j}}
\sum\limits_{m=1}^{M}
~\sum\limits_{\bm k\in\{-K,\dots,K-1\}^{d-1}}
\varphi_{m}\hat{\Psi}_{\bm j,\bm k}\\
&\cdot \rme^{-2\pi \rmi \left(\bm j\cdot\bm{\mu}_{m,\bm k}\right)/cM}\,,
\end{aligned}
\end{align}
where the entries in $\bm \mu_{m,\bm k}$ are the nearest integers to $\bm x_m+\bm k$. 
Here we have used the abbreviations
\begin{align*}
\begin{aligned}
\Psi_{\bm j,\bm k} &:=\,\prod\limits_{i=1}^{d-1}\Psi(2\pi \bm j/N_x)\,,\\
\hat\Psi_{\bm j,\bm k} &:=\,\prod\limits_{i=1}^{d-1}\left(\frac{c}{\sqrt{2\pi}}\right)\hat{\Psi}((\bm x_{m})_i-(\bm\mu_{m,\bm k})_i)\,,
\end{aligned}
\end{align*}
for the needed evaluations of $\Psi$ and $\hat{\Psi}$.

Further remarks on the implementation of the NED- and NER-NUFFT, 
as well as a summary about the properties of the Kaiser-Bessel function 
and its Fourier transform can be found in \cite{Fou03,HalSchZan09b}.

\section{The Experimental Setup}
\label{sec:experimental}

Before we turn to the evaluation of the algorithm we describe the photoacoustic 
setup. A detailed explanation and characterization of the working principles of our setup
can be found in \cite{ZhaLauBea08}. It consists of a  Fabry P\'{e}rot (FP) polymer film
sensor for interrogation \cite{Bea05,BeaPerMil99}, a $50\,\mathrm{Hz}$ pulsed laser source and a subsequent optical parametric 
oscillator (OPO) which emits optical pulses. These pulses have a very narrow 
bandwidth and can be tuned within the visible and near infrared range. 
The optical pulses propagate through an optical fiber. 
When the light is emitted it diverges and impinges upon a sample. 
Some of this light is absorbed and partially converted into heat.
This leads to a pressure rise generating a photoacoustic wave, which is then recorded via the FP-sensor head.
The sensor head consists of an approximately $\mathrm{38\,\mu m}$ thick polymer
(Parylene C) which is sandwiched between two dichroic dielectric coatings.
These dichroic mirrors have a noteworthy transmission characteristic.
Light from $600$ to $1200\,\mathrm{nm}$ can pass the mirrors largely
unattenuated, whereas the reflectivity from $1500$ to $1650\,\mathrm{nm}$
(sensor interrogation band) is about 95\% \cite{ZhaLauBea08}. 
The acoustic pressure of the incident photoacoustic wave produces a change in the optical
thickness of the polymer film. A focused continuous wave laser, operating
within the interrogation band, can now determine the change of thickness
at the interrogation point via FP-interferometry. 
The frequency response of this specific setup of up to $100\,\mathrm{MHz}$ has been
analytically predicted, based on a model used in \cite{BeaPerMil99} and experimentally confirmed \cite{ZhaLauBea08}.
There is a linear roll-off reaching zero at $57.9\,\mathrm{MHz}$, with a subsequent rise.

\section{Comparison of the NER-NUFFT reconstruction with FFT and time reversal}
\label{sec:comp1_NER_NUFFT}

In this section several reconstruction methods will be compared for regular grids. 
This will be done with synthetic data as well as with experimental data. 
All CPU based reconstructions are carried out with a workstation PC (Quad Core @ 3.6 GHz).
All parameters that are not unique to the reconstruction method are left equal.

\subsection{Synthetic data} 
\label{sec:synthetic_data}

\begin{figure}
	\begin{center}\includegraphics[width=0.45\columnwidth]{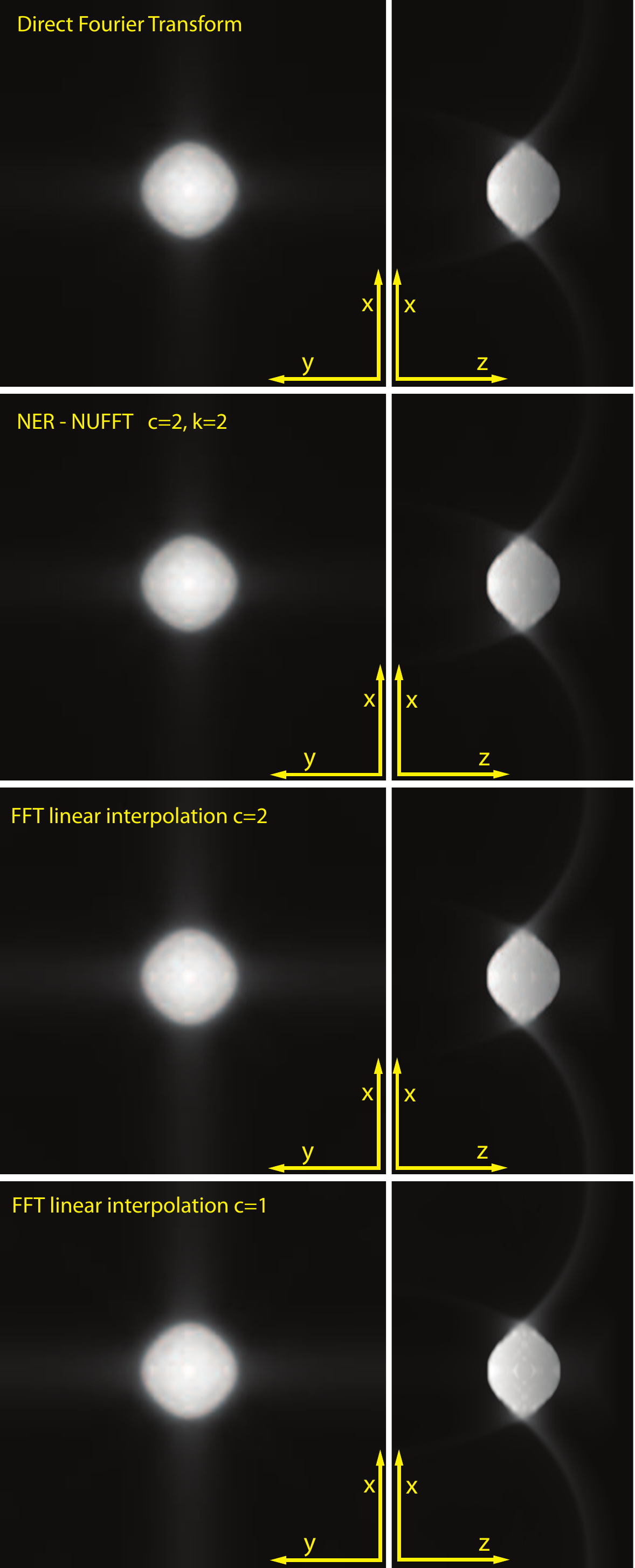}\end{center}

	\caption{Maximum intensity projections (MIPs) in the $xy$ and $xz$ plane of different reconstructions, for a solid sphere. The direct Fourier Transform (top) serves as ground truth.
$c$ denominates the upsampling factor in the time domain and $k$ the interpolation width of the Kaiser-Bessel function in \eqref{eq:equiv_rec}.
	\label{fig:BallComparison}}
	\end{figure}

For the comparison of different implementations of the FFT based reconstruction we conduct a forward simulation 
of a solid sphere on a $200\cdot 200\cdot 100$ computational grid, using the \emph{k-wave 1.1} Matlab library. 
The maximum intensity projections of the $xy$ and the $xz$ plane of all reconstructions are shown in figure \ref{fig:BallComparison}.

To obtain the closest possible numerical reconstruction of the Fourier based inversion formula, we directly evaluate the right hand side 
of formula \eqref{eq:equiv_rec}, and subsequently invert a 3D equispaced Fourier transform by applying the conventional 3D (inverse) FFT.
We call this reconstruction \emph{direct FT}. It serves as ground truth for computing the correlation coefficient (appendix \ref{App:Qualitmeas}).

For the NUFFT reconstruction, the temporal frequency oversampling factor in \eqref{eq:NER_NUFFT}  is set to $c=2$ and the interpolation length is set to $K=2$.
In the linear interpolation FFT case, we use both $c=1$ and $c=2$. The FFT-reconstruction with $c=1$ was conducted via the \emph{k-wave} toolbox, which doesn't provide oversampling options out of the box. However the oversampling still can be achieved in a computationally not optimal way by adding zeros at the end of the data term in the time dimension. After reconstruction, the temporal dimension translates into the $z$ axis. In the $xy$ dimension no oversampling is performed.

The correlation coefficient and the computational time of the methods can be found in table \ref{tab:BallComparison}. 
The errors indicate the superiority of the NUFFT reconstruction in comparison to linear interpolation, with a comparable computational effort (Table \ref{tab:BallComparison}).
The results also show that in the FFT case, an artificial oversampling in the temporal frequency dimension is highly recommended.

\begin{table}
\begin{center}%
\begin{tabular}{c|c|c|c|c|c}
 & time & \multicolumn{4}{c}{(correlation-100) in \%}\tabularnewline
 & (s) &  3D & $xy$ & $xz$ & $yz$\tabularnewline
\hline 
NER (c=2) & 59 & 0.005 & 0.0003 & 0.001 & 0.001\tabularnewline
FFT (c=2) & 56 & 3.457 & 0.54 & 0.45 & 0.45\tabularnewline
FFT (c=1) & 53 & 14.00 & 0.65 & 0.81 & 0.81\tabularnewline
\end{tabular}\end{center}
\caption{The first column compares computational times. In the last four columns the 
difference to a full correlation with the direct FT reconstruction method is given in \%, for the 3D data and the 3 maximum intensity projections.}
\label{tab:BallComparison}
\end{table}

\subsection{Experimentally acquired data} 
\label{sec:ChickComparisonNER-NUFFT}
For an overall qualitative assessment two data sets, of a 3.5, and a 5 day old chick embryo,  are used \cite{LiuMauHerZabSan14}.
This corresponds to the development stages HH21 and HH27 of the Hamburger \& Hamilton (HH) criterion \cite{HamHam92}.
The data are sampled with a spatial step size of  $60\,\mathrm{\mu m}$, covering an area of $1.008\cdot1.008 \mathrm{mm^2}$ (3.5 days) and $1.02\cdot 1.02\mathrm{mm^2}$ (5 days) and a time step size of $16\, \mathrm{ns}$, corresponding to a maximum frequency of 31.25 MHz. To avoid aliasing the signal is low pass filtered to the maximal spatial frequency of 25.3 MHz.
A full reconstruction with the NER-NUFFT for the 5 day old embryo is shown in figure \ref{fig:ChickenOrganFull}.

The time reversal reconstruction is performed via the \emph{k-wave} toolbox. The spatial upsampling factor in the $x$, $y$ direction is set to 2.
For time reversal this is realized by linearly interpolating the sensor data to a finer grid, whereas for the FFT based reconstructions zero padding in the Fourier domain is performed.

The oversampling factor for the FFT reconstructions in the time domain is $c=2$. The number of time steps used for the reconstruction
covers more than twice the depth range of the visible objects and is 280 for the HH21 and 320 for the HH27 embryo.

In table \ref{tab:ComparisonChickenTime} a comparison for the computational time is
shown. For the NUFFT case, $\Psi$ as defined in \eqref{eq:PSI} can be precomputed, which roughly halves reconstruction time in subsequent reconstructions using the same discretization, as has been already reported in \cite{SchZanHolMeyHan11}.
Moreover, the computation time improves by a factor of 200 when using FFT-based reconstructions instead of time reversal.

\begin{table}
\begin{center}
\begin{tabular}{c|c|c}
\multicolumn{1}{c}{} & \multicolumn{2}{c}{time (s)}\tabularnewline
embryo stage \cite{HamHam92} & HH21 & HH27\tabularnewline
\hline 
NER-NUFFT (k=2,c=2) & 21 & 24\tabularnewline
NER-NUFFT with precomputed $\Psi$ & 13 & 14\tabularnewline
FFT with linear interpolation (c=2) & 20 & 23\tabularnewline
Time Reversal & 7236 & 7659\tabularnewline
\end{tabular}
\end{center}
%
%
\caption{Comparison of the computational effort of two chick embryo data sets, with different reconstruction methods.}
\label{tab:ComparisonChickenTime}
\end{table}

\begin{figure}
\begin{center}\includegraphics[width=0.5\columnwidth]{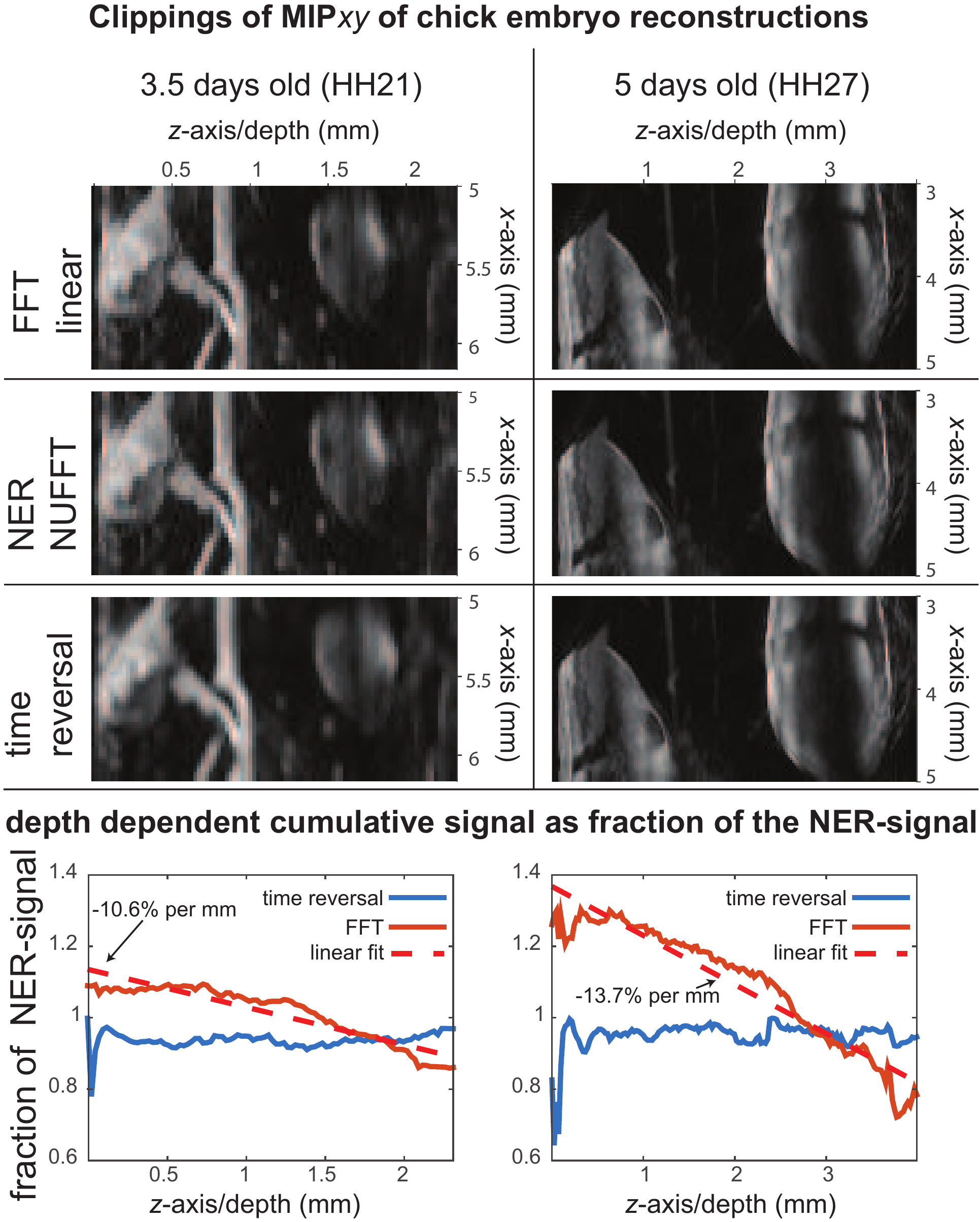}\end{center}

\caption{The top shows clippings of the MIPs along the $y$-axis for the two chick embryos. The three
reconstruction methods from top to bottom are: FFT with linear interpolation, NER-NUFFT and time reversal.
All reconstructions are normalized, so their maximum value is 1. On the bottom graphs, the cumulative signal for each $z$-axis 
layer is plotted as a fraction of the corresponding NER-NUFFT layer. While the depth dependent signal between the time reversal
reconstruction and the NER-NUFFT roughly remains the same, for the FFT with linear interpolation a fall-off can be observed. 
The linear fit suggests a reduction of 10.6 \% per mm for the 3.5 days old chick embryo and 13.7 \% per mm for the 5 days old data.
\label{fig:IntensityFallOff}}
\end{figure}

The relative Tenenbaum sharpness (appendix \ref{App:Qualitmeas}) for the 3D data
of the 3.5 days old chick embryo was slightly better for the NER-NUFFT reconstruction (44.2)
than for time reversal (43.0) and FFT with linear interpolation (41.1). 
A comparison of clippings of the maximum intensity projection (MIP) in the $xz$ plane is shown
in figure \ref{fig:IntensityFallOff}. The time reversal reconstruction seems smoothed compared to the FFT reconstructions, which is probably
a result of the different spatial upsampling modalities.

In the bottom graphs of figure \ref{fig:IntensityFallOff} the cumulative reconstructed signal 
for each layer is plotted, as fraction of the NER-NUFFT cumulative signal.
The additional fall-off for the FFT with linear interpolation has been determined by a line fit.
For the 3.5 day old embryo it was 10.6 \% per mm and 13.7 \% per mm for the 5 day old embryo. 
While it intuitively makes sense that the $z$-axis is primarily affected by errors
introduced by a sub-optimal implementation of equation \ref{eq:exact},
this problem needs further research to be fully understood.

\section{Non-Equispaced Sensor Placement}\label{sec:Non-Equispaced-Sensor-Placement}

The current setups allow data acquisition at just one single sensor point for each laser pulse 
excitation. Since our laser is operating at $50\,\mathrm{Hz}$ data recording of a typical 
sample requires several minutes. Reducing this acquisition time is a crucial step in advancing
photoacoustic tomography towards clinical and preclinical application. 
Therefore, we try to maximize the image quality for a given number of acquisition points and a given region of interest.

Our newly implemented NEDNER-NUFFT is ideal for dealing with non-equispaced positioned sensors,
as error analyses for the NED- and the NER-NUFFT indicate \cite{Fou03}.
This newly gained flexibility of sensor positioning offers many possibilities
to enhance the image quality compared to a rectangular grid.

Also any non-equispaced grids that may arise from a specific experimental setup can be efficiently computed
via the NEDNER-NUFFT approach.

\begin{figure}[tbh]
\begin{center}\includegraphics[width=0.45\columnwidth]{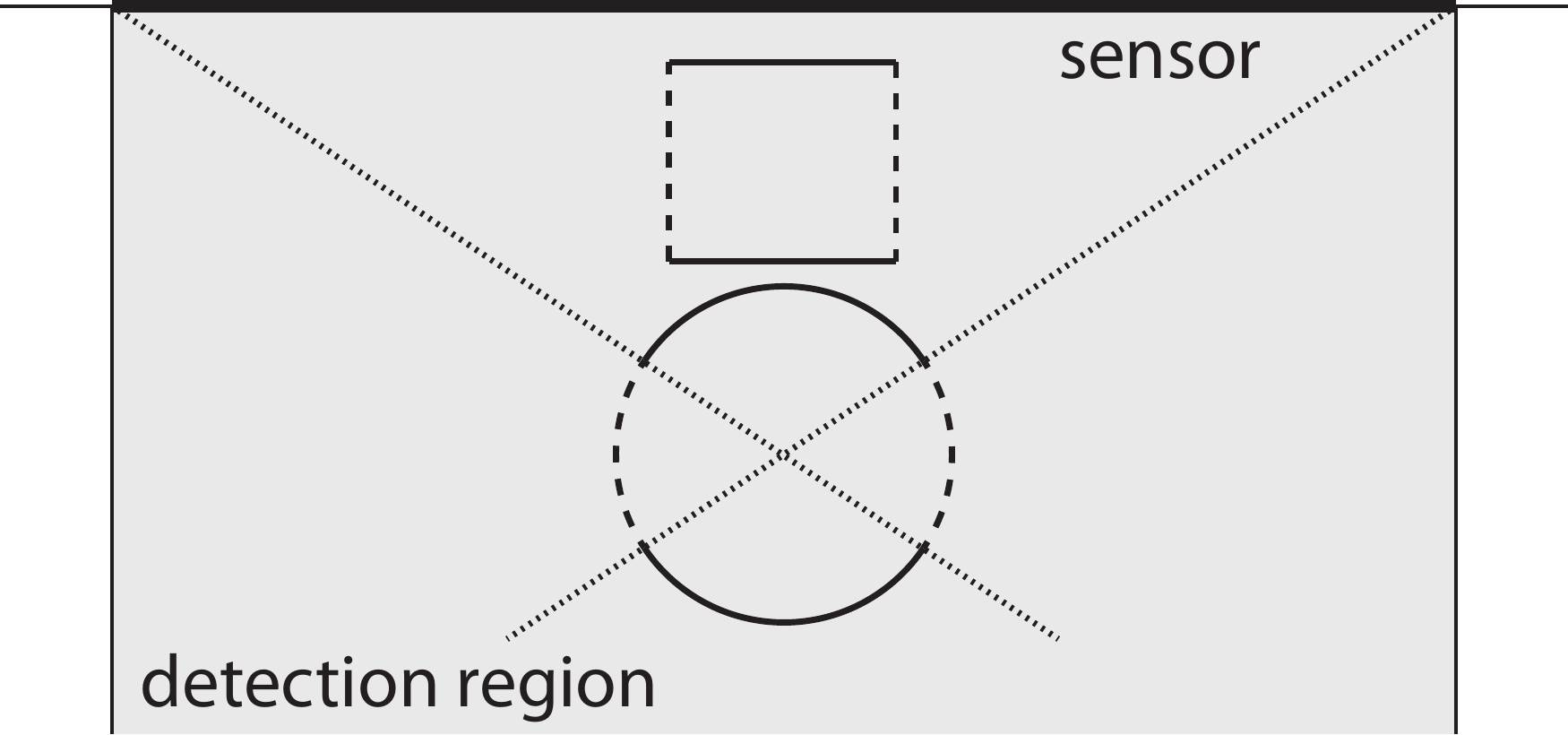}\end{center}

\caption{Depiction of the limited view problem. Edges whose normal vectors cannot
intersect with the sensor surface are invisible to the sensor. The
invisible edges are the coarsely dotted lines. The detection region
is marked by a grey background. The finely dotted lines are used to
construct the invisible edges. Edges perpendicular to the sensor surface
are invisible for a plane sensor. \label{fig:Fieldofview}}
\end{figure}

\subsection{Equi-angular and equi-steradian projections} 
\label{sec:sensor_mask}
In this article, we use the NEDNER-NUFFT to tackle the limited view or limited aperture problem, for the case of a limited number of available detectors, which can be placed discretionary on a planar surface.
To understand the limited view problem, it is helpful to define a detection region.
According to \cite{XuWanAmbKuc04}, this is the region which is enclosed
by the normal lines from the edges of the sensor. 
Mathematically speaking, the wave front propagates on straight lines in the direction of the singularity \cite[Chapter VIII]{Hoe03}. As a consequence the reconstruction is locally stable if the straight line through the normal to the object boundary passes through the detector surface \cite{LouQui00}. Therefore certain edges are invisible to the
detector, as depicted in figure \ref{fig:Fieldofview}. 
One approach to overcome this problem experimentally has been made by enclosing the target in a reverberant cavity \cite{CoxArrBea07}.
In addition, a lot of effort has been made to enhance reconstruction techniques in order to deal with the limited view problem 
\cite{FriQui15, DanTaoLiuWan12,XuWanAmbKuc04,TaoLiu10, AnaWanZhaKruReiKru08, WanSidAnaOraPan11}. 

Our approach to deal with this problem is different. It takes into account that in many cases the limiting factor is the number of sensor
points and the limited view a consequence of this constraint.
We use an irregular grid arrangement that is dense close to a center of interest 
and becomes sparser the further away the sampling points are located.
We realize this by means of an equi-angular, or equi-steradian sensor arrangement,
where for a given point of interest each unit angle or steradian
gets assigned one sensor point. This arrangement can also be seen as a mock hemispherical detector.

For the equi-angular sensor arrangement a point of interest is chosen.
Each line, connecting a sensor point with the point of interest, encloses
a fixed angle to its adjacent line. In this sense we mimic a circular
sensor array on a straight line. The position of the sensor points
is pictured on top of the third image in figure \ref{fig:tree}. 

The obvious expansion of an equi-angular projection to 3D is the 
equi-steradian projection. Here we face a problem analogous to the problem of placing
equispaced points on a 3D sphere and then projecting the points, from
the center of the sphere, onto a 2D plane outside
the sphere (the detector plane). 

The algorithm used for this projection is explained in detail in appendix \ref{App:equi-ster}. Our input variables
are the diameter of the detection region, which we define as the diameter of the disc where the sensor points are located,  
the distance of the center of interest from the sensor plane $r$ and the desired number of acquisition points.
In the top left section of figures \ref{fig:ChickenSecondFeat} and \ref{fig:ChickenOrganComp} the sensor arrangements are depicted.

\subsection{Weighting term} \label{sec:weighting}

To determine the weighting term $h_m$ in \eqref{eq:disc_rec} for 3D we introduce a function
that describes the density of equidistant points per unit area $\rho_{p}$. 
In our specific case, $\rho_{p}$ describes the density on a sphere around a center of interest.
Further we assume that $\rho_{p}$ is spherically symmetric and decreases quadratically
with the distance from the center of interest $r$: $\rho_{p,s}\propto1/r^{2}$.
We now define $\rho_{p,m}$ for a plane positioned at distance
$r_{0}$ from the center of interest. In this case $\rho_{p,s}(r)$
attenuates by a factor of $\sin\alpha$, where $\alpha=\arcsin(r_{0}/r)$
is the angle of incidence. Hence $\rho_{p,m}\propto r_{0}/r^{3}$.
This yields a weighting term of:
\[h_{m}(r)\propto r^{3}\]

Analogously we can derive $h_m$ for 2D:
\[h_{m}(r)\propto r^{2}\]

For the application of this method to the FP setup it is noteworthy
that there is a frequency dependency on sensitivity which itself depends on the angle of incidence.
These characteristics have been extensively discussed in \cite{CoxBea07}.

\section{Application of the NEDNER-NUFFT with Synthetic Data in 2D}  
\label{sec:2D Comparisons}

\begin{figure}[tbh]
\begin{center}\includegraphics[width=0.45\columnwidth]{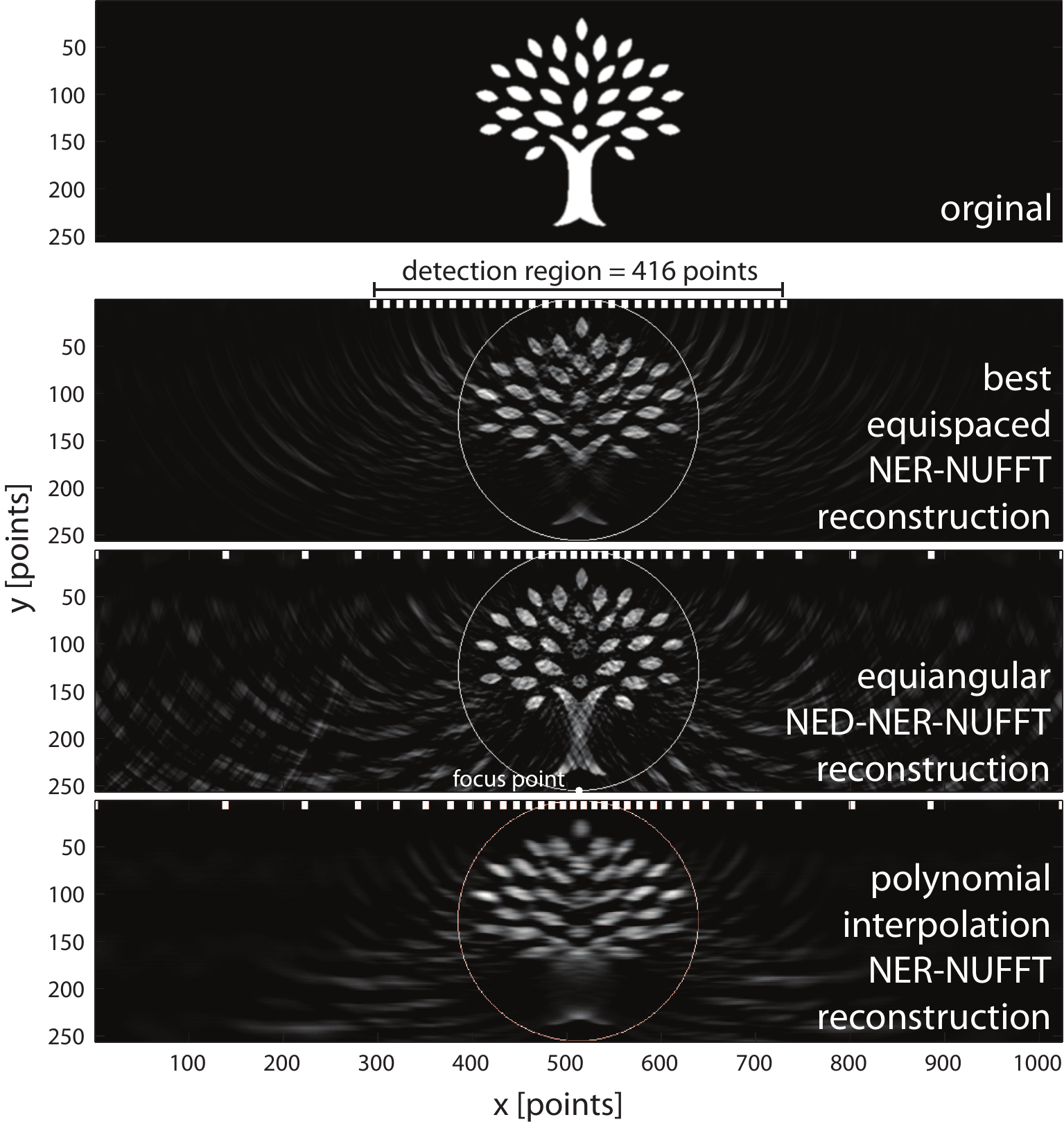}\end{center}
\caption{Various reconstructions of a tree phantom (top) with different sensor
arrangements. All sensor arrangements are confined to 32 sensor points.
The sensor positions are indicated as white rectangles on the top
of the images. The second image shows the best (see figure \ref{fig:Tree_CC})
equispaced sensor arrangement, with a distance of 13 points between
each sensor. The third image shows the NEDNER-NUFFT reconstruction
with equi-angular arranged sensor positions. The bottom image shows
the same sensor arrangement, but all omitted sensor points are polynomially
interpolated and afterwards a NER-NUFFT reconstruction was conducted.
\label{fig:tree}}
\end{figure}

A tree phantom, designed by Brian Hurshman and licensed under CC BY
3.0%
\footnote{http://thenounproject.com/term/tree/16622/%
}, is chosen for the 2 dimensional computational experiments on a grid
with $x=1024$ $z=256$ points. A forward simulation is conducted
via \emph{k-wave 1.1} \cite{TreCox10}. The forward simulation of the k-wave
toolbox is based on a first order k-space model. A PML (perfectly
matched layer) of 64 grid points is added.
Also white noise is added to obtain an SNR (signal to noise ratio) of $30\,\mathrm{dB}$.

In figure \ref{fig:tree} our computational phantom is shown at the
top. For each reconstruction a subset of 32 out of the 1024 possible
sensor positions was chosen. In figure \ref{fig:tree} their positions
are marked at the top of each reconstructed image. For the equispaced
sensor arrangements, we let the distance between two adjacent sensor
points sweep from 1 to 32, corresponding to a detection region sweep from 32 to 1024. The sensor points are always centered
in the $x$-axis.

To compare the different reconstruction methods we use the correlation coefficient
and the Tenenbaum sharpness. These quality measures are explained in appendix \ref{App:Qualitmeas}.

\begin{figure}
\begin{center}\includegraphics[width=0.45\columnwidth]{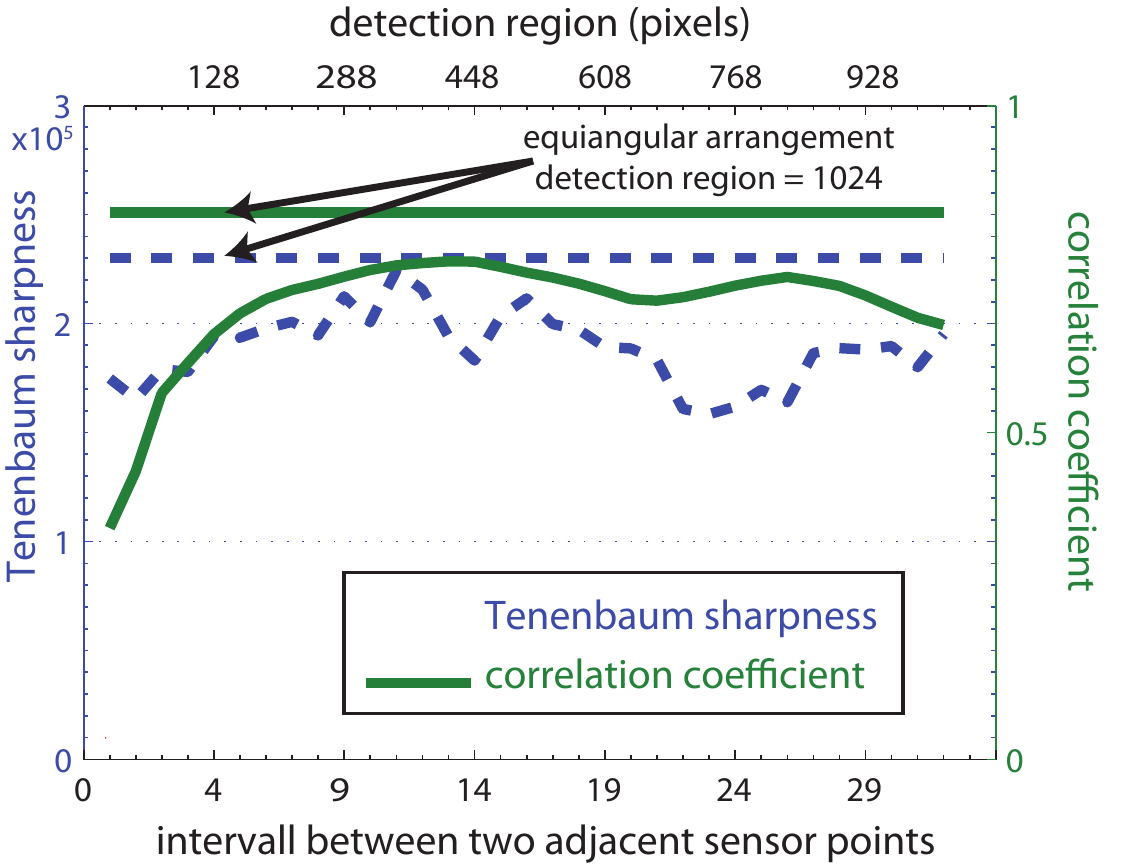}\end{center}
\caption{Correlation coefficient and Tenenbaum sharpness for equispaced sensor
arrangements with intervals between the sensor points reaching from
1 to 32. The maximum of the correlation coefficient is at 13. The
corresponding reconstruction is shown in figure \ref{fig:tree}. The
straight lines indicate the results for the equi-angular projection.\label{fig:Tree_CC}}
\end{figure}

We apply the correlation coefficient only within
the region of interest marked by the white circle in figure \ref{fig:tree}.
The Tenenbaum sharpness was calculated on the smallest rectangle,
containing all pixels within the circle. The results are shown in
figure \ref{fig:Tree_CC}. 

The Tenenbaum sharpness for the equi-angular sensor placement is
23001, which is above all values for the equispaced arrangements.
The correlation coefficient is 0.913 compared to 0.849, for the best
equispaced arrangement. In other words, the equi-angular arrangement
is 42.3 \% closer to a full correlation than any equispaced grid.

In figure \ref{fig:tree} the competing reconstructions are compared.
While the crown of the tree is depicted quite well for the equispaced
reconstruction, the trunk of the tree is barely visible. This is owed
to the limited view of the detection region. 
As the equispaced interval and the detection region increase,
the trunk becomes visible, but at the cost of the crown's quality.
In the equi-angular arrangement a trade off between these two effects
is achieved. Additionally the weighting term for the outmost sensors
is 17 times the weighting term for the sensor point closest to the
middle. This amplifies the occurrence of artifacts, particularly
outside the region of interest. 

The bottom image in figure \ref{fig:tree} shows the equi-angular
sensor arrangement, where the missing sensor points are polynomially interpolated to an equispaced grid
and a NER-NUFFT reconstruction is applied afterwards. The interpolation is conducted for every time step
from our subset to all 1024 sensor points. The correlation coefficient
for this outcome was 0.772 while the sharpness measure is 15654.
This outcome exemplifies the clear superiority of the NUFFT to conventional
FFT reconstruction when dealing with irregular grids.

\section{Application of the NEDNER-NUFFT with Experimental Data in 3D} 
\label{sec:NEDNER-realdata}

We will now examine if the positive effects of the NEDNER-NUFFT reconstruction with non-equispaced detectors are transferred 
to 3D data. For these comparisons we use the data sets of the two chick embryos already presented in
section \ref{sec:ChickComparisonNER-NUFFT}. By the use of polynomial interpolation for each time step, 
we map this data to discretionarily placed points on the acquisition plane. 
Thus sensor data is obtained for regular and irregular grids with arbitrary step sizes. 
The sensor positions are indicated by the red dots in figures  \ref{fig:ChickenSecondFeat} and \ref{fig:ChickenOrganComp}.

\begin{figure}
\begin{center}\includegraphics[width=0.45\columnwidth]{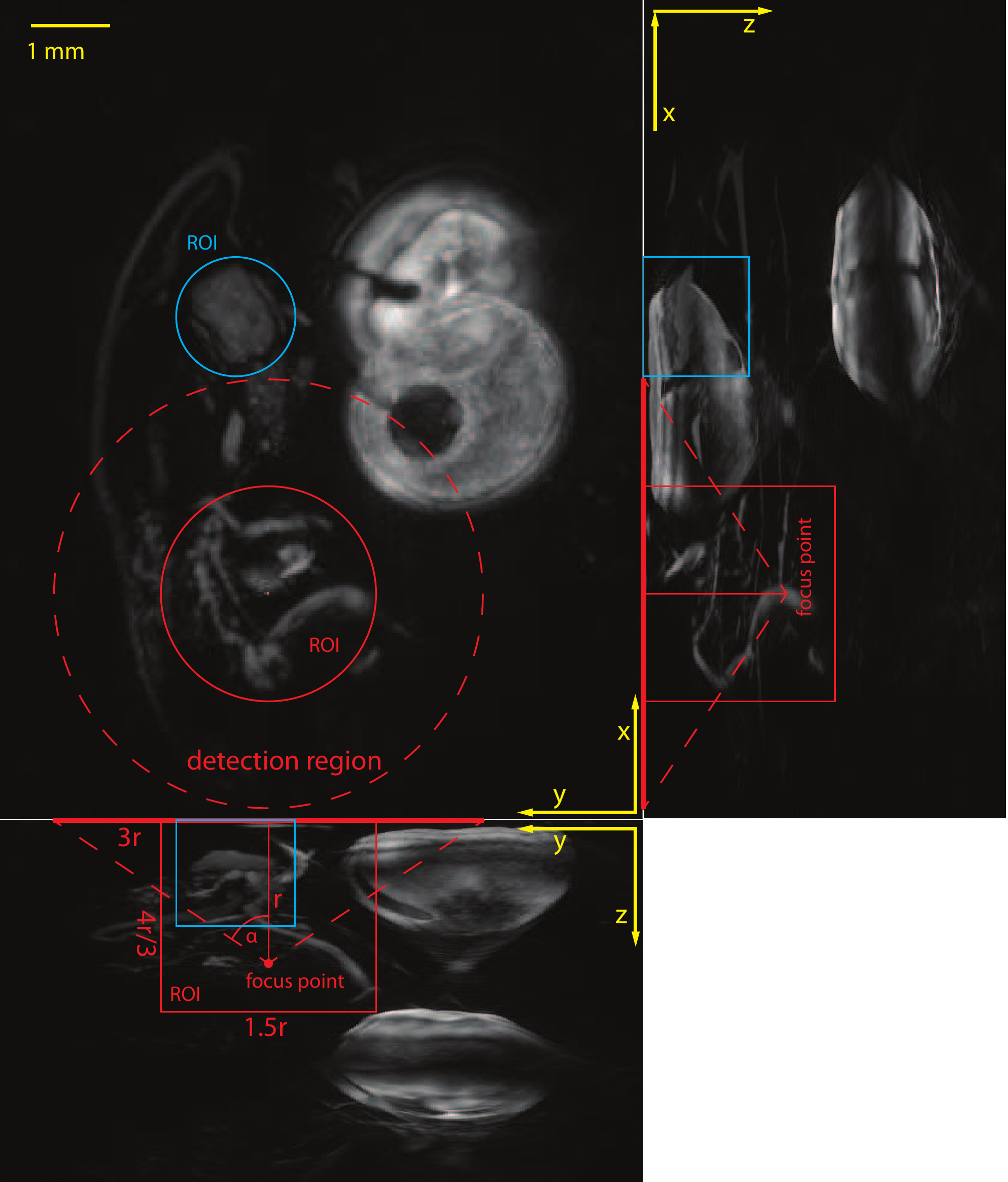}\end{center}
\caption{MIPs (maximum intensity projections) of a full NER-NUFFT reconstruction
of a 5 day old chick embryo (HH27), cropped along the $y$-axis. Two cylindrical ROIs (regions of interest) are indicated, each by a circle and two rectangles.
The red ROI is discussed in figure \ref{fig:ChickenOrganComp}, the blue ROI in figure \ref{fig:ChickenSecondFeat}.
For the red ROI the focus point for the equi-steradian arrangement is shown and the detection region, which marks the area where sensor points are located. 
The distance $r$ of the 'focus point' from the plane governs the size of the detection region
and the ROI, according to the proportions shown.
Within the detection region all sensor points occupy the same steradian from the focus point's perspective. 
\label{fig:ChickenOrganFull}
}
\end{figure}

This procedure allows us to use a full reconstruction as a ground truth
and thus ensures a quantitative quality control via the correlation coefficient (Appendix \ref{App:Qualitmeas}).
We are also safe from any experimental errors that could be introduced between measurements.
A drawback is that we can only interpolate to step sizes $\geqq 60 \mathrm{\mu m}$ without loss of information.
Therefore the presented images are always made with rather few sensor points and naturally of a lower quality.
However we want to emphasize that this is a result of our experimental procedure.

For all comparison reconstructions the NEDNER-NUFFT has been used for practical reasons.
While the NER-NUFFT cannot deal with non rectangular grids, it is equivalent to the NEDNER-NUFFT 
for rectangular regular grids. Using the NEDNER-NUFFT the spacing of the computational grid can be chosen freely. 
It corresponds to the width of the Kaiser-Bessel function for interpolation (see \eqref{eq:NED_NUFFT}).
If the computational grid is much finer than the local sensor point density, a strong signal close to the sensor surface
will produce high intensity spots with an intensity distribution according to the Kaiser-Bessel function, instead of a homogeneous area.
Making the computational grid coarser than the sensor point density 
produces a more blurry reconstruction with a reduced lateral resolution.
The computational grid therefore is chosen as fine as possible without reducing the lateral resolution.

We use the two chick embryo data sets to extract the irregular sensor data via layer-wise polynomial interpolation.
A clipping of the MIP$xz$  of reconstruction of both chick embryos is shown in figure \ref{fig:IntensityFallOff}.
A full NER-NUFFT reconstruction of the 5 day old chick embryo is shown in figure \ref{fig:ChickenOrganFull}. 
For the comparisons we define a region of interest (ROI) in the form of a cylinder with a height to diameter ratio of 8:9.
The area where the sensor points are located for a given reconstruction will be called the \emph{detection region}.
The proportions between the ROI and the detection region is the same for all measurements as depicted in \ref{fig:ChickenOrganComp}.

In order to avoid spatial aliasing the time data have been low pass filtered with a cut off frequency according to $F_{cutoff}=c_{sound}/2dx$ where $c_{sound}$ is the sound speed and $dx$ the step size. In the equi-steradian grid, the (locally varying) stepsize $dx$ has been defined as the distance to the nearest neighboring point. 

\begin{figure}
\begin{center}\includegraphics[width=0.5\columnwidth]{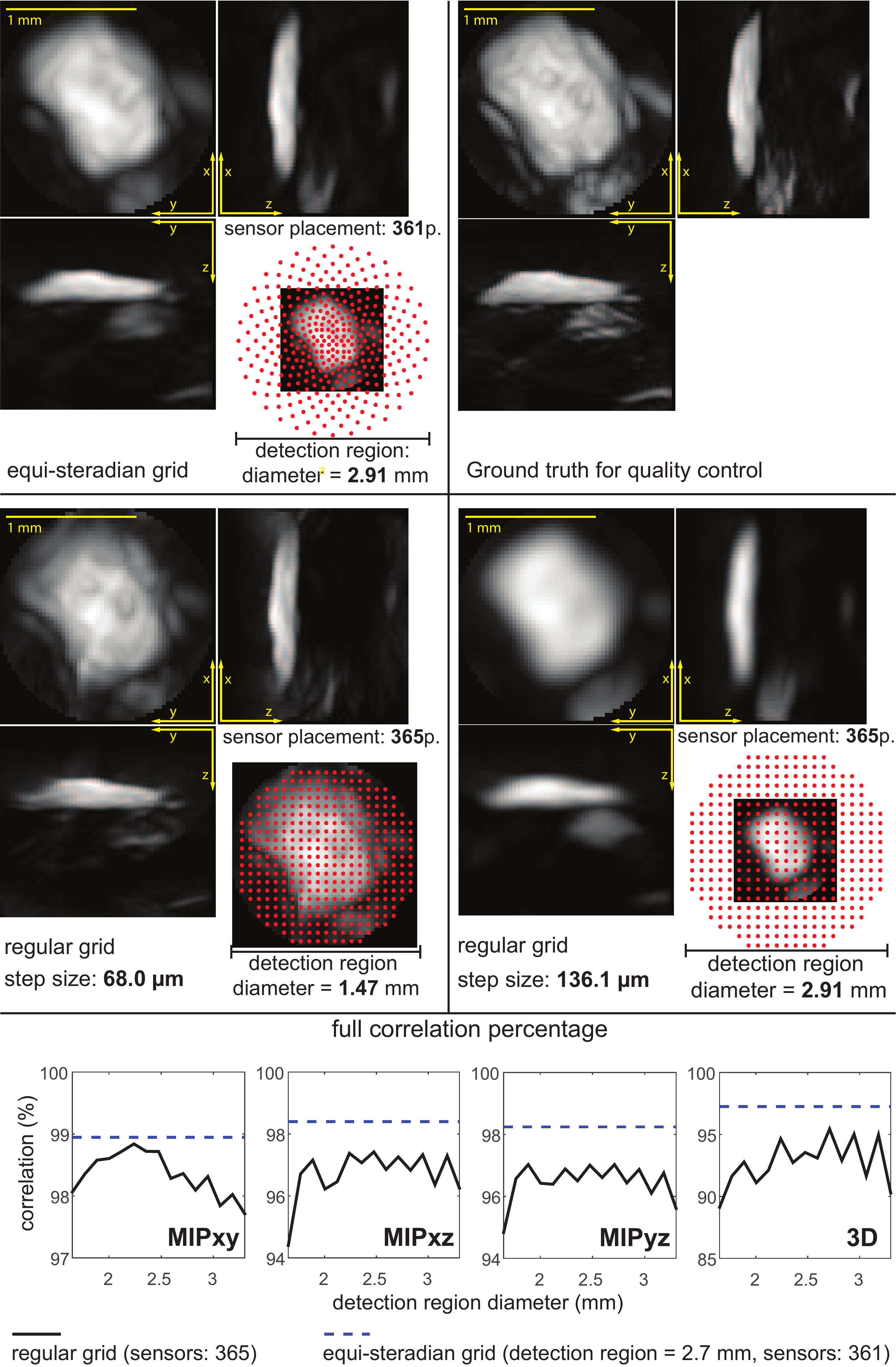}\end{center}
\caption{Comparisons of different reconstructions for a region of interest (ROI), (marked blue in figure \ref{fig:ChickenOrganFull}) with roughly 360 sensor points.  All reconstructions are presented in the form of MIPs. The top right shows the ground truth, the top left the equi-steradian sensor arrangement. Below them the reconstructions of the regular grids for the largest and the smallest detection region are depicted. On the bottom the correlation coefficient for different regular grids is shown for the three MIPs and the 3D-data.
\label{fig:ChickenSecondFeat}
}
\end{figure}

\begin{figure*}
\begin{center}\includegraphics[width=1\columnwidth]{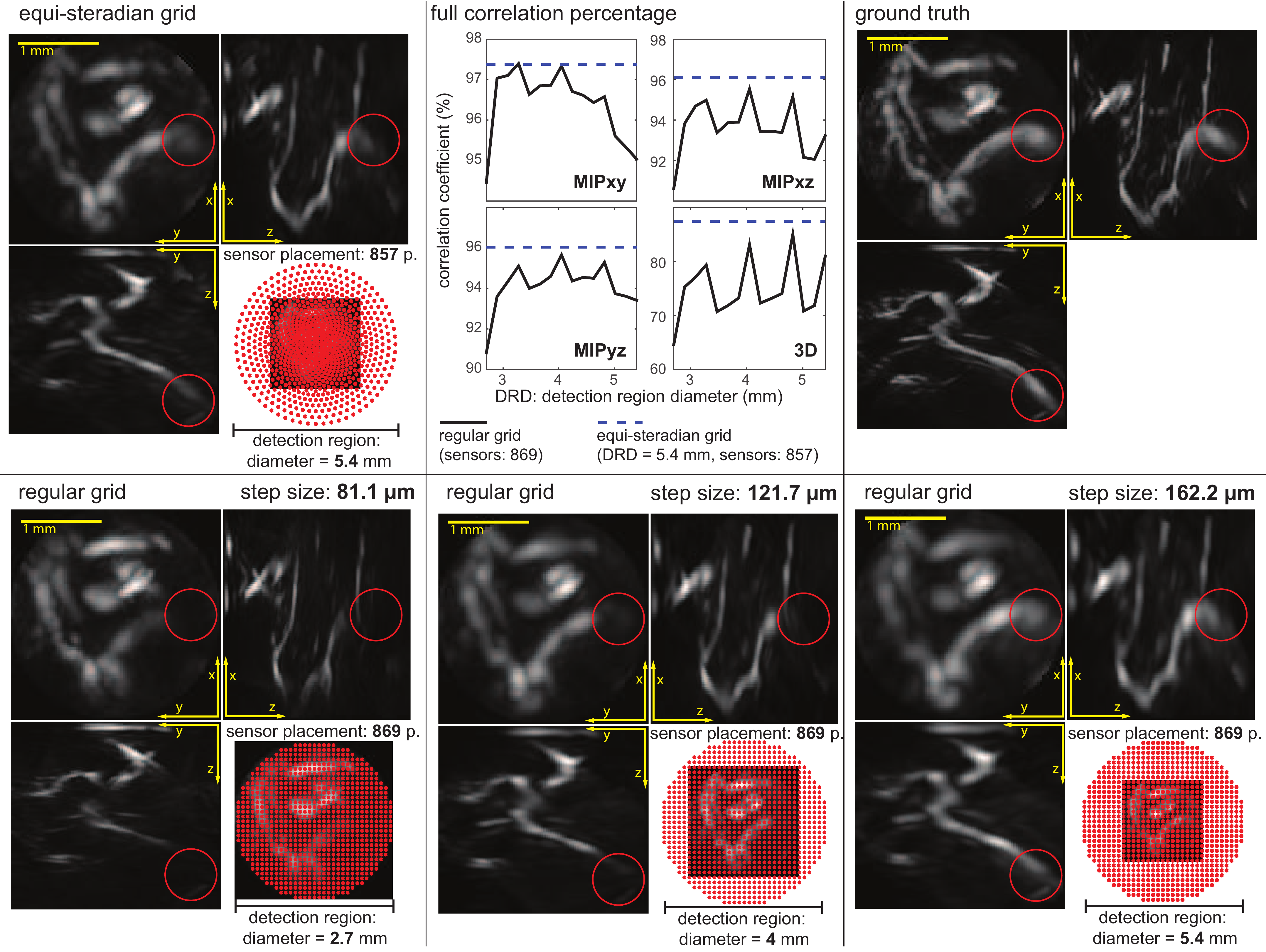}\end{center}
\caption{Comparisons of different reconstructions for a region of interest (ROI), (marked as red in figure \ref{fig:ChickenOrganFull}) with roughly 870 sensor points.
All reconstructions have been cropped to the cylindrical ROI. 
Three maximum intensity projections (MIPs), are shown for every reconstruction. The sensor placement is indicated by red dots, which overlay 
the MIP$xy$ on the bottom right of the dedicated segment. The three bottom segments show regularly 
arranged sensor points. 
The detection region diameter (DRD) for the three bottom segments is 2.7, 4 and 5.4 mm while the number of sensor points always remains 869.
Certain features (red circle) are not visible for the small detection regions, due to the limited view problem.
As the detection region becomes larger, these features start to appear, at the cost of overall resolution.
The equi-steradian arrangement shown on the top left still shows these features, while maintaining a high resolution.
On the top right a reconstruction with all original points of the detection region is shown. This was used as ground truth.
On the top center segment, the correlation coefficient, for different regular grids is shown. \label{fig:ChickenOrganComp}}
\end{figure*}

\begin{figure}\label{fig:ChickenTRI}
\begin{center}\includegraphics[width=0.5\columnwidth]{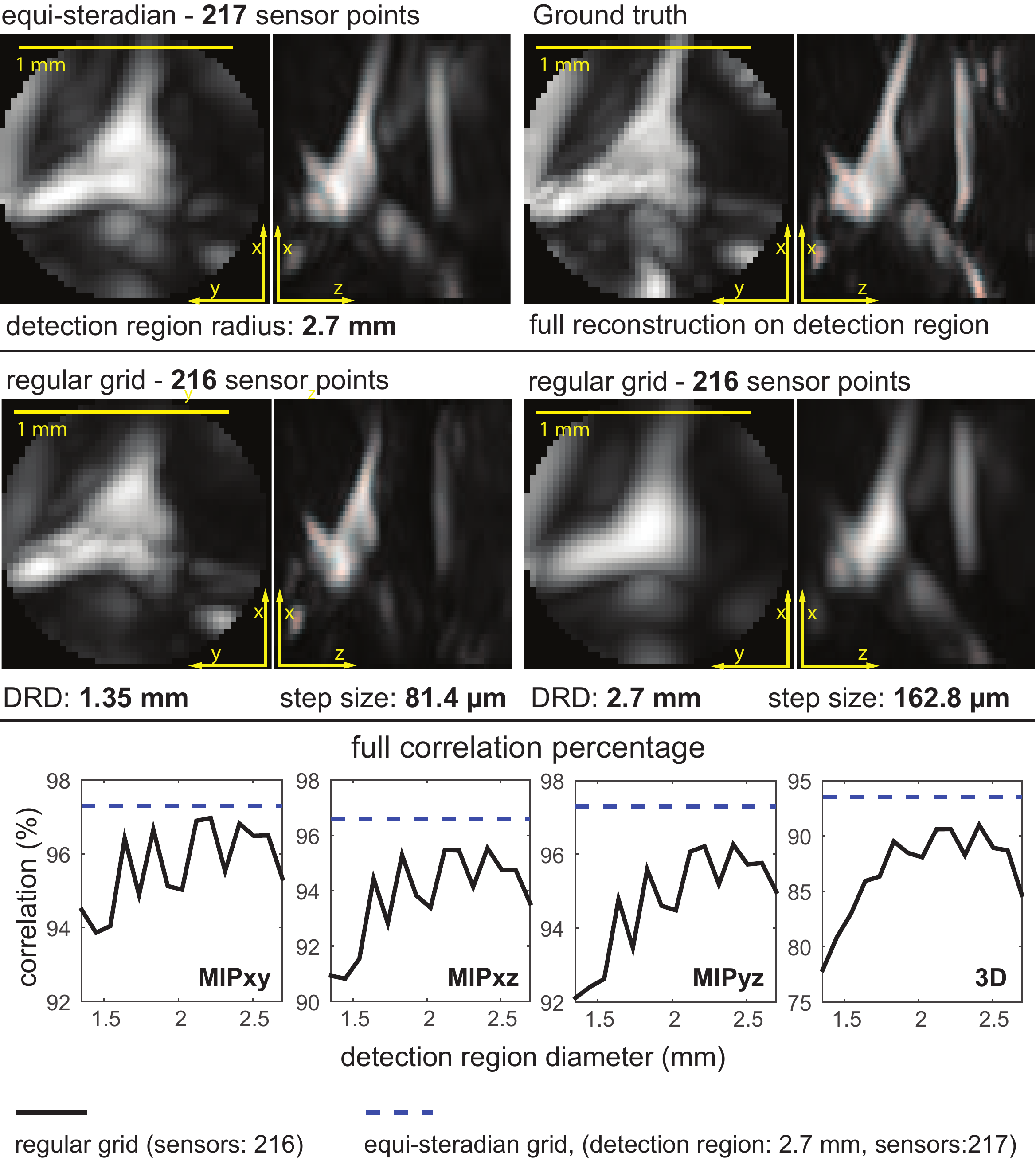}\end{center}
\caption{Reconstruction comparisons of the center region of the 3.5 day old chick embryo (HH21). The depiction is analog to figure \ref{fig:ChickenSecondFeat}, without 
the MIP$yz$ and the sensor placement.}
\end{figure}

We now conduct a fair comparison between the equi-steradian sensor arrangement described in appendix \ref{App:equi-ster}
and regular grid arrangements for the given ROI. This is done by maximizing image fidelity, while always using (approximately) the same
number of sensor points. 
The comparisons are undertaken for three different ROIs. This is done to show that 
the advantages of the equi-steradian arrangement are not confined to a single case,
but rather consistent for different features and volume sizes.
All selected ROIs need to have a detection region, that is fully covered by the underlying data set.

The results are shown in figures \ref{fig:ChickenSecondFeat}, \ref{fig:ChickenOrganComp} and \ref{fig:ChickenTRI}.
The figures are organized in a similar manner and depict different reconstructions via maximum intensity projections (MIPs).
On the top left segment, the equi-steradian grid arrangement is shown. The top right segment shows the ground truth: A NEDNER-NUFFT reconstruction using all original sensor points that are placed within the detection region.
In the segments below, the regular grid reconstructions are shown. On the left the detection region coincides with the region of interest, on the right it is coincides with the detection region.
Figures \ref{fig:ChickenSecondFeat} and \ref{fig:ChickenOrganComp} also show the sensor point placement.
While in the small detection region the reconstructions have a good resolution, edges are blurred and certain features are invisible. 
As the detection region increases, these features appear, at the cost of reduced overall resolution.
The equi-steradian grid arrangement has a rather high resolution towards the center, while still displaying the mentioned features.

All the above figures contain four graphs, which depict the correlation coefficient for the three MIPs and the volume data. The detection region for the regular grids
is increased and shown on the $x$-axis, while the number of acquisition points stays constant. The computational grid was chosen 
to be as small as possible but greater than the step size and has been increased in steps of $30 \mathrm{\mu m}$ in order to have a consistent
$\mathrm{\mu m/pixel}$ spacing in the reconstruction. 

The results show that the equi-steradian arrangement consistently produces reconstructions that outperform every regular grid arrangement.
It provides a good combination of a large detection region and high resolution at the region of interest.
This is demonstrated by the juxtaposition of different reconstructions and also confirmed via the correlation coefficient.

\section{Conclusions}\label{sec:results}
\subsection{Summary and results}

We computationally implemented a 3D non-uniform FFT
photoacoustic image reconstruction, called NER-NUFFT (non equispaced range-non
uniform FFT) to efficiently deal with the non-equispaced Fourier transform evaluations
arising in the reconstruction formula.

In the computational results, it could be shown that the NER-NUFFT is 
much closer (more than 100 times in the test piece) to perfect correlation than the FFT reconstruction
with linear interpolation.

We then used real data sets for comparison, recorded with a FP-planar sensor setup \cite{ZhaLauBea08}
and included the \emph{k-wave} time reversal algorithm.
Regarding reconstruction time the results of \cite{SchZanHolMeyHan11} and \cite{HalSchZan09b} could be confirmed for 3D, where the
FFT with linear interpolation performs similar to NER-NUFFT. Additionally, 
the NER-NUFFT reconstruction time could be significantly reduced (almost halved), 
if the values of the interpolation functions $\Psi$ and $\hat \Psi$ had already been pre-computed
for the chosen discretization.
The time reversal computation took more than 300 times longer on a CPU, than any FFT based reconstruction.
Concerning image quality, the NER-NUFFT and time reversal reconstruction
perform on a very similar level, while the conventional FFT method fails to correctly image the depth-dependent intensity fall-off.
 While this fall-off is almost synchronous for time reversal and NER-NUFFT, there was an additional 
intensity drop of about 10 \% per mm in the linear interpolation FFT based reconstruction.

The second application of the NUFFT approach concerned the applicability of
irregular grid arrangements, which were new in photoacoustic tomography.
In fact, this was done by implementing the NEDNER-NUFFT  (non equispaced data-NER-NUFFT).
Our goal was to maximize image quality in a given region of interest, using a limited 
number of sensor points. To do this we developed an equi-angular sensor placement
for 2D and an equi-steradian placement in 3D,
which assigns one sensor point to each angle/steradian for a given
center of interest.

For the 2D simulations we showed that this 
arrangement enhances the image quality for a given region of interest
and a confined number of sensor points in comparison to regular grids.

In 3D we used the aforementioned chick embryo data and 
reconstructed with an interpolated subset of the original sensor data.
We thus conducted a fair comparison between regular grid arrangements 
and the equi-steradian arrangement with a limited number of sensor points
for three regions of interest. 
While the volume of these regions ranged from 1.7 to 13.7 $\mathrm{mm^3}$
the shape always remained a cylinder with a height to diameter ratio of 8:9. 

For our regions of interest, the correlation of the equi-steradian arrangement 
to the full reconstruction, was consistently higher than any regular
grid arrangement, using an almost equal number of sensor points.

\subsection{Discussion}

For the case of regular sampled grids the results of \cite{HalSchZan09b} where confirmed
for 3D in the synthetic data experiments. The synthetic data results further
show the importance of using a zero-pad factor of at least 2 in the time domain,
when using FFT based reconstruction methods. In the case of real data,
The main identifiable difference was the additional intensity drop for greater depth of the FFT reconstruction
in comparison to the other two methods. 
The great computational advantage of using FFT based reconstructions
makes it the most suitable method for most cases in a planar sensor geometry setup.
There was no detectable difference in the reconstruction quality between the NER-NUFFT and time reversal.
From our point of view,
the use of the NER-NUFFT therefore seems to be especially useful in the case of high resolution imaging
in relatively deep-lying regions.
%
%

The NEDNER-NUFFT implementation allowed to efficiently reconstruct data from
non-equispaced sensor points. This is used to extend the primary application of a planar sensor surface, 
recording images over a large area, by the possibility to image a well defined region of interest with a shorter acquisition time.
Thus we designed a sensor mask to better image small regions in larger depths 
at a fixed number of sensor points, using a design that projects an equispaced hemispherical detector geometry onto a planar sensor surface.
For this case our sensor arrangement produced consistently better
reconstructions than any regular grid, because it allowed to maintain a high resolution
within our region of interest, while still
capturing features that could only be detected
outside the region of interest.
In our test examples, the NEDNER-NUFFT further enhances the image quality in deep regions while maintaining
a reasonable computational effort.

In comparison to a real hemispherical detector, there is an increase of acoustic attenuation.
This is countered by greater accessibility, scalability and flexibility on the planar detector.
The region of interest not necessarily needs to fit into a spherical shape, the size of the region of interest just has an upper bound and the
number of acquisition points is limited by the measurement time. 

As an outlook, we mention that the case where the field of view is much larger than 
the imaging depth has not been investigated in this paper.
For this case a similar approach of expanding the field of view by
non-equispaced sensor point placement is possible. This could mitigate 
the image degradation towards the boundaries of the detection region.
However the achievable benefit of such a method would
decrease with an increase of the ratio of the detection region area to the
detection region boundary and the maximum imaging depth.

%
%

\section*{Acknowledgment}
This work is supported by the Medical University of Vienna, the 
European projects FAMOS (FP7 ICT 317744) and FUN OCT (FP7 HEALTH 201880), 
Macular Vision Research Foundation (MVRF, USA), Austrian Science Fund (FWF), Project P26687-N25 
(Interdisciplinary Coupled Physics Imaging), 
and the Christian Doppler Society 
(Christian Doppler Laboratory "Laser development and their application in medicine").
We further want to thank Barbara Maurer and Wolfgang J. Weninger from the Center for Anatomy and Cell Biology at the Medical University of Vienna
for providing us with the chick embryo.

\bibliographystyle{plain}
\bibliography{\BibPath strings,\BibPath articles,\BibPath books,\BibPath infmath,\BibPath infmath_books,\BibPath infmath_report,\BibPath infmath_talks,\BibPath infmath_theses,\BibPath inproceedings,\BibPath preprints,\BibPath proceedings,\BibPath theses,\BibPath unsubmitted,\BibPath websites}



\appendix
\subsection{Appendix: Algorithm for equi-steradian sensor arrangement}
\label{App:equi-ster}

In our algorithm, the diameter of the 
detection region and the distance of the center of
interest from the sensor plane is defined. The number of sensor points
$N$ will be rounded to the next convenient value.

Our point of interest is placed at $z=r_{0}$, centered at a square
$xy$ grid. The point of interest is the center of a spherical coordinate
system, with the polar angle $\theta=0$ at the $z$-axis towards
the $xy$-grid. 

First we determine the steradian $\Omega$ of the spherical cap from
the point of interest, that projects onto the acquisition point plane via 
\[\Omega=2\pi\left(1-\cos\left(\theta_{max}\right)\right)\;.\]
This leads to a unit
steradian $\omega=\Omega/N$ with $N$ being the number of sensors
one would like to record the signal with. The sphere cap is then subdivided
into slices $k$ which satisfy the condition 
\[\omega\, j_{k}=2\pi\left(\cos\left(\theta_{k-1}\right)-\cos\left(\theta_{k}\right)\right)\,,\]
where $\theta_1$ encloses exactly one unit steradian $\omega$ and $j_k$ has to be a power of two, in order to guarantee some symmetry. 
The value of $j_k$ doubles, when $r_s>1.8\cdot r_k$, where $r_s$ is the chord length between two points on $k$ and $r_k$ is the distance to the closest point on $k-1$.
These values are chosen in order to approximate local equidistance between acquisition points on the sensor surface.

The azimuthal angles for a slice $k$ are calculated according to:
\[\varphi_{i,k}=\left(2\pi i\right)/j_{k}+\pi/j_{k}+\varphi_{r}\,,\]
with $i=0,\ldots,j-1$, where 
\[\varphi_{r}=\varphi_{j_{k-1},k-1}+\left(k-1\right)2\pi/(j_{k-1})\]
stems from the former slice $k-1$ . 
The sensor points are now placed on the $xy$-plane at the position indicated
by the spherical angular coordinates:
\[\left(\mathrm{pol,az}\right)=\left(\left(\theta_{k}+\theta_{k+1}\right)/2,\varphi_{i,k}\right)\]

\subsection{Appendix: Quality measures}
\label{App:Qualitmeas}

In the case where a ground truth image is available, we choose the correlation coefficient $\rho$, which is a 
measure of the linear dependence between two images $U_{1}$ and $U_{2}$.
Its range is $\left[-1,1\right]$. A correlation coefficient close
to 1 indicates linear dependence \cite{Sch11}. It is defined via
the variance $\mbox{Var}\left(U_{i}\right)$ of each image and the
covariance $\mbox{Cov}\left(U_{1},U_{2}\right)$ of the two images:
\begin{equation}\label{eq:corrcoeff}
\begin{array}{cl}
\rho\left(U_{1},U_{2}\right)=\frac{\mbox{Cov}\left(U_{1},U_{2}\right)}{\sqrt{\mbox{Var}\left(U_{1}\right)\mbox{Var}\left(U_{2}\right)}}\;.
\end{array}
\end{equation}

We decided not to use the widely applied $L^{p}$ distance measure because
it is a morphological distance, meaning it defines the distance
between two images by the distance between their level sets. Therefore
two identical linearly dependent images can have a correlation coefficient
of $1$ and still a huge $L^{p}$ distance. 
This can be dealt with by normalizing the data as in \cite{XuWanAmbKuc04,TaoLiu10}.
We choose the correlation coefficient instead,
because in experimentally acquired data single high intensity artifacts can occur, 
which would have a disproportionately large effect on the normalized $L^{p}$ distance. 

In case of experimentally collected data,
there are only a few methods available
for the comparison of different reconstruction methods.
A possible way for measuring sharpness is obtained from a measure for the high frequency content
of the image \cite{GroYouLig85}. 
Out of the plethora of published focus functions we select the Tenenbaum
function, because of its robustness to noise:

\begin{equation}
\begin{array}{cl}
F_\text{Tenenbaum}=\underset{x,y}{\sum}\left(g*U_{x,y}\right)^{2}+\left(g^{T}*U_{x,y}\right)^{2}\,,
\end{array}
\end{equation}

with $g$ as the Sobel operator: 
\begin{equation}
g=\left(\begin{array}{ccc}
-1 & 0 & 1\\
-2 & 0 & 2\\
-1 & 0 & 1
\end{array}\right)\;.
\end{equation}

Like the $L^{2}$ norm and unlike the correlation coefficient, the
Tenenbaum function is an extensive measure, meaning it increases with
image dimensions. Therefore we normalized it to $\overline{F}_\text{Tenenbaum}=F_\text{Tenenbaum}/N$,
where $N$ is the number of elements in $U$.

\end{document}